\documentclass{ws-ijmpd}
\usepackage[super,compress]{cite}
\usepackage{lineno,hyperref}
\usepackage{amsmath}
\usepackage{amsfonts}
\usepackage{amssymb}
\usepackage{siunitx}
\usepackage{multirow}
\usepackage{amsmath}
\modulolinenumbers[5]

\usepackage{bigints}
\usepackage{lipsum}

\begin{document}
	
	\markboth{Tuñacao, Abac}
	{Modified White Dwarfs from GUP}
	
	\catchline{}{}{}{}{}
	%
	%%%%%%%%%%%%%%%%%%%%%%%%%%%%%%%%%%%%%%%%%%%%%%%%%%%%%%%%%%%%%%%%%%%%
	
%	\title{Equations of State and Mass-Radius Relations of Quadratic Generalized Uncertainty Principle-modified White Dwarfs with Arbitrary Temperatures%\footnote{For the title, try not to use more than 3 lines.
		%Typeset the title in 10~pt Times roman, uppercase and boldface.}  
%	}

\title{Finite Temperature Considerations in the Structure of Quadratic GUP-modified White Dwarfs}
	
	\author{JAMES DAVID M. TUÑACAO%\footnote{Typeset names in
		%8~pt roman, uppercase. Use the footnote to indicate the
		%present or permanent address of the author.}
	}
	
	\address{Department of Physics, University of San Carlos, Nasipit, Talamban\\
		Cebu City, Cebu 6000, Philippines\\
		%\footnote{State completely without abbreviations, the
		%affiliation and mailing address, including country. Typeset in
		%8~pt Times italic.}\\
		jamesdtunacao@gmail.com}
	
	\author{ADRIAN G. ABAC}
	
	\address{Max Planck Institute for Gravitational Physics (Albert Einstein Institute), Am Mühlenberg 1, D-14476 Potsdam, Germany\\
		adrian.abac@aei.mpg.de}
	\address{Department of Physics, University of San Carlos, Nasipit, Talamban\\
		Cebu City, Cebu 6000, Philippines\\}
	
	\author{ROLAND EMERITO S. OTADOY}
	
	\address{Department of Physics, University of San Carlos, Nasipit, Talamban\\
		Cebu City, Cebu 6000, Philippines\\
		rsotadoy@usc.edu.ph}
	
	\maketitle
	
	\begin{history}
		\received{Day Month Year}
		\revised{Day Month Year}
	\end{history}

\begin{abstract}
In quantum gravity phenomenology, the effect of the generalized uncertainty principle (GUP) on white dwarfs has been given much attention in the literature. However, these studies assume a zero temperature equation of state (EoS), consequently excluding young white dwarfs whose initial temperatures are substantially high. To that cause, this paper calculates the Chandrasekhar EoS and resulting mass-radius relations of finite temperature white dwarfs modified by the quadratic GUP, an approach that extends Heisenberg's uncertainty principle by a quadratic term in momenta. The EoS was first approximated by treating the quadratic GUP parameter as perturbative, causing the EoS to exhibit expected thermal deviations at low pressures, and conflicting behaviors at high pressures, depending on the order of approximation. We then proceeded with a full numerical simulation of the modified EoS, and showed that in general, finite temperatures cause the EoS at low pressures to soften, while the quadratic GUP stiffens the EoS at high pressures. This modified EoS was then applied to the Tolman-Oppenheimer-Volkoff equations and its classical approximation to obtain the modified mass-radius relations for general relativistic and Newtonian white dwarfs. The relations for both cases were found to exhibit the expected thermal deviations at small masses, where low-mass white dwarfs are shifted to the high-mass regime at large radii, while high-mass white dwarfs acquire larger masses, beyond the Chandrasekhar limit. Additionally, we find that for sufficiently large values of the GUP parameter and temperature, we obtain mass-radius relations that are completely removed from the ideal case, as high-mass deviations due to GUP and low-mass deviations due to temperature are no longer mutually exclusive.
\end{abstract}

	
	\ccode{PACS numbers:}
	
	%\tableofcontents
	
	\section{Introduction}
	\label{section: Introduction}
	
	Several theories associated with quantum gravity such as string theory \cite{amati,blau2009string, gross1988string, konishi1990minimum}, path integral quantum gravity \cite{padmanabhan1985physical, padmanabhan1986role, padmanabhan1987limitations, greensite1991there}, loop quantum gravity \cite{garay1995quantum,hossenfelder2013minimal}, and doubly special relativity (DSR) \cite{cortes,ali2009discreteness, ali2011minimal} predict the existence of a minimum measurable length, on the order of the Planck length $l_{p}\left(\sim 10^{-35} \mathrm{~m}\right.$). In string theory for example, it is suggested that there is a minimum possible distance at which strings interact, that being the length of the string itself \cite{amati}. Additionally, in black hole physics \cite{maggiore,scardigli1999generalized, maggiore1994quantum}, particularly in the observation of photons scattered through Hawking radiation \cite{hawking1975particle}, it is also suggested that a photon cannot carry information more detailed than its own wavelength \cite{maggiore}. The existence of such a minimal length introduces a modification to the Heisenberg uncertainty principle (HUP), which is then called the generalized uncertainty principle (GUP), with the modification believed to have a gravitational origin \cite{tawfik, garay1995quantum, hossenfelder2013minimal}.

An approach to GUP, consistent with string theory and black hole physics, proposes a modification of the HUP at the Planck scale called the \textit{quadratic} GUP - quadratic in the sense that the HUP is extended by an extra term of momentum squared such that $\Delta x \Delta p \sim \hbar [1 + \beta (\Delta p)^2]$, which leads to the modified phase space volume $(1 + \beta p^2)^{-3} d^3xd^3p $ \cite{chang2002effect}. The constant coefficient $\beta$ is known as the quadratic GUP parameter, whose exact value is debated by several previous studies \cite{wang, brau2006minimal, das2008universality, scardigli2017gup, das2021bounds, tamburini2022constraining}. Another approach, consistent with string theory, black hole physics, and DSR is called the linear GUP, with modified uncertainty principle $\Delta x \Delta p \sim \hbar (1 - \alpha \Delta p)$ and phase space volume $(1 - \alpha p)^{-4} d^3xd^3p $ \cite{ali2011minimal}. Numerous other approaches exist in the literature, such as the linear-quadratic GUP \cite{belfaqih2021white}, various higher-order GUP approaches \cite{nouicer2007quantum,pedram2012higher1,pedram2012higher2,
shababi2017two,chung2018new,chung2019new,hassanabadi2019analysis,
shababi2020new,petruzziello2021generalized,du2022new}, and extended GUP for nonzero cosmological constant \cite{ong2018generalized}. For simplicity, this paper focuses on the quadratic GUP alone.

%paramanik2022generalized

The effect of GUP in various physical systems is an active area of research in quantum gravity phenomenology \cite{herkenhoff2023framework, merriam2022thermodynamics, carvalho2022gravitational, anacleto2022generalized, khosropour2022statistical, rani2022casimir, shababi2022effects, hamil2022gup, abac,abac2021implications,anacleto2021quasinormal, bosso2021quantum, bosso2020quantum, twagirayezu2020generalized}. In particular, References \citen{rashidi,ong,mathew} note that GUP effects are most evident in systems of ultra-high energy, strong gravity, or ultra-fine length scales. Compact objects such as white dwarfs therefore arise as a natural candidate for investigation. Most notable is the effect of GUP on the maximum allowable mass of a white dwarf, known classically as the Chandrasekhar mass limit ($\sim 1.4M_{\odot}$) \cite{chandrasekhar}. Under normal circumstances, a white dwarf surpassing this limit is expected to undergo gravitational collapse, evolving into denser compact objects; a neutron star for objects less than $3M_{\odot}$, or a black hole for much more massive objects. \cite{bally2006birth}. When applying quadratic GUP effects, curiously this isn't the case:

Wang, Yang, and Zhang in Ref. \citen{wang} considered an approximate GUP modification to white dwarfs composed of an ultra-relativistic Fermi gas, and found that the Chandrasekhar mass limit is increased by a small positive factor, inferring that the quadratic GUP tends to resist the collapse of white dwarfs. Moussa in Ref. \refcite{moussa} also approximated the GUP modification to find that the white dwarf radius tends toward infinity as the stellar mass approaches the Chandrasekhar limit. Rashidi in Ref. \refcite{rashidi} employed the same assumptions as Wang et al. without approximating the GUP modification, and found that as the central Fermi momentum of a white dwarf increases, the stellar mass and radius increase without bounds, essentially removing the Chandrasekhar mass limit. The same findings are reported by Ong in Ref. \refcite{ong}, where heuristic calculations of GUP corrections to relativistic Fermi gases resulted in white dwarfs becoming unbounded in size. In the same paper, Ong suggests that this is avoided if $\beta$ is a negative value. Finally, a complete investigation into the removal of the mass limit is employed by Mathew and Nandy in Ref. \refcite{mathew} by deriving the analytical form of the GUP-modified Newtonian structure equations for white dwarfs. Their mass-radius relations still showed that the masses and radii increase without bounds, but the relations may terminate at certain finite masses beyond Chandrasekhar's limit, if the white dwarf's central momentum is restricted by either the deformation of phase space due to GUP, or neutronization of white dwarf matter due to inverse $\beta$-decay.

It is important to note however, that the above studies were performed under the mathematical idealization of a cold white dwarf, by employing a zero temperature Equation of State (EoS). Using the finite temperature EoS would prove to be more realistic for new-born white dwarfs that start out at high temperatures, before cooling in the billions of years after. A review of existing literature suggests this has not been done before. Additionally, these studies limited their investigations by only studying quadratic GUP effects in classical white dwarfs, by using the stellar structure equations of Newtonian gravity \cite{wang, rashidi, mathew}. While this assumption is sufficient in studying the essential features of white dwarfs without GUP \cite{boshkayev}, recall that GUP effects appear at ultra-high energies, which are found in the most massive white dwarfs. These objects have gravities that significantly warp spacetime, hence, studying GUP effects in the context of general relativity (GR) should be more appropriate.

To this cause, we present the objective of this paper, which is to investigate the effects of the quadratic GUP on the structure of finite temperature white dwarfs, particularly by obtaining the modified mass-radius relations in Newtonian gravity and GR. The white dwarf is assumed to be a plasma ball of pressure-ionized matter, whose degeneracy pressure and total energy density are related by the Chandrasekhar EoS \cite{boshkayev, carvalho2018general}. GUP modifications appear primarily in the EoS, which are then fed to the stellar structure equations to obtain the mass-radius relation. We find that in both Newtonian gravity and GR, hotter white dwarfs in the low-mass regime acquire slightly larger masses, yet our consideration of thermal effects does not protect the white dwarf from surpassing the Chandrasekhar mass limit. The value of $\beta$ also determines how strongly the modified relations deviate from the ideal case, with larger $\beta$ aftecting more low-mass white dwarfs and vice versa. Furthermore, for sufficiently large values of $\beta$, deviations due to GUP and temperature overlap, producing a mass-radius relation that is completely removed from the ideal case.

For completeness, this study derives approximate and exact forms for the modified finite temperature EoS, both employing an energy dispersion relation describing Fermi gases of arbitrary relativity, as opposed to limiting our consideration to ultra-relativistic gases. Furthermore, the structure equations are solved numerically for a variety of white dwarf central pressures, allowing us to investigate GUP effects over a wide range of stellar masses and radii (on the mass-radius relation), as opposed to restricting investigations on the white dwarf's limiting mass alone.

This paper is structured as follows. In Section \ref{section: section_2}, we derive the modified thermodynamic properties of white dwarf matter to arrive at the modified Chandrasekhar EoS. In Section \ref{section: section_3}, we explore two approaches to calculating this EoS: an approximate solution that involves expanding the EoS as a Taylor series, which we will call the perturbative approach, and a full numerical approach that involves re-expressing the EoS in terms of energies instead of momenta, which we refer to as the non-perturbative approach. In Section \ref{section: section_4}, we introduce the Tolman-Oppenheimer-Volkoff (TOV) structure equations in GR, the corresponding Newtonian approximation, and feed the non-perturbative EoS into the equations to obtain the mass-radius relations. Finally, in Section \ref{section: section_5}, we discuss our conclusions, recommendations, and possible extensions to the study.

	\section{Quadratic GUP and the equation of state}
	\label{section: section_2}
	
		\subsection{The quadratic GUP}
		
		\label{subsection:quadratic_GUP}
		
		In the introduction, we mentioned that the quadratic GUP is an approach consistent to string theory and black hole physics. The following examples illustrate this point: Ref. \refcite{amati} analyzed the ultra high-energy scatterings of strings (in string theory), to check the inconsistencies of quantum gravity at the Planck scale, finding that strings can't interact at distances shorter than its own length $\lambda_{s}=\sqrt{\hbar \alpha}$, where $\alpha$ is the string's tension \cite{tawfik, amati, scardigli1999generalized, gross1988string}. Meanwhile, thought experiments proposed to measure a black hole’s apparent horizon also lead to a generalization of the HUP that agrees with the above suggestion from string theory \cite{maggiore1993generalized}. This measurement is performed by observing the photons scattered by a black hole emitting Hawking radiation, where detecting said radiation allows us to make a black hole “image”.  Both theories suggest a commutaton relation \cite{tawfik,kempf1995hilbert,chang2002effect} of the form:
\begin{equation}
[\mathbf{x},\mathbf{p}] = i\hbar(1+\beta\mathbf{p}^2) \label{eq:kempf}
\end{equation}
as introduced in the seminal paper by Kempf, Mangano, and Mann (1995) (Ref. \citen{kempf1995hilbert}). From \eqref{eq:kempf}, we can derive the uncertainty relation
\begin{equation}
\Delta x \Delta p \geq \frac{\hbar}{2}\left(1+\beta\left\langle p^{2}\right\rangle\right) \rightarrow \frac{\hbar}{2}\left[1+\beta(\Delta p)^{2}+\beta\langle p\rangle^{2}\right] \label{eq:delXdelP_quadratic GUP}
\end{equation}
Where we have used the relation $(\Delta p)^{2}=\left\langle p^{2}\right\rangle-\langle p\rangle^{2}$ on the RHS. The parameter $\beta$ is given by
\begin{equation}
\beta = \frac{\beta_0}{M_p^2 c^2} = \frac{\beta_{0} l_{p}^{2}}{\hbar^{2}}
\end{equation}
where $M_p$ is the Planck mass, $c$ is the speed of light, and $\beta_0$ is a dimensionless GUP parameter.
The exact value of $\beta_{0}$ is currently unknown, but previous works have attempted to provide estimates for its boundaries. Wang, Yang, and Zhang (2010) proposed a lower bound of $\beta_{0}>10^{4}$, imposed by the Hagedorn temperature of relativistic strings \cite{wang}. Brau and Buisseret (2006) proposed an upper bound of $\beta_{0}<10^{34}$, obtained by comparing the energy spectrum of the gravitational quantum well modified by a first order perturbation of $\beta$, with energy spectrum results obtained from the GRANIT experiment \cite{brau2006minimal}. Das and Vagenas (2008) propose various upper bounds to $\beta_{0}$, obtained by showing that the existence of a minimal length produces quantum gravity corrections to various quantum phenomena \cite{das2008universality}: (1) $\beta_{0}<10^{36}$ from the accuracy in precision measurements of the Lamb shift of the H atom, (2) $\beta_{0}<10^{50}$ from the accuracy of direct measurements of Landau levels using STM (scanning tunneling microscopy), and (3) $\beta_{0}<10^{21}$, the value needed for a GUP induced current (in the quantum tunneling of electrons in STM) to add up to the charge of just one electron, in the span of 1 year. Scardigli, Lambiase, Vagenas (2016) also propose that $\beta=82 \pi/5 \left(\beta_{0} \sim 10^{10}\right)$, from computing the Hawking temperature for a Swarzschild black hole \cite{scardigli2017gup}.

Recently, A. Das, S. Das, Mansour, and Vagenas (2021) obtained upper bounds by comparing graviton and photon speeds in a GUP-modified curved spacetime with speeds obtained from gravitational wave events GW150914 and GW190521. Considering GUP modifications in graviton speed only, an upper bound of $\beta_{0}< 2.56 \times 10^{60}$ was obtained, while considering modifications in both graviton and photon speed generated an upper bound as small as $\beta_{0}< 8.83 \times 10^{35}$ \cite{das2021bounds}. Tamburini, Feleppa, and Thide (2022) obtained an upper bound by comparing the orbital angular momentum acquired after light is lensed by a GUP-modified rotating black hole, with experimental data acquired for M87. They found an upper bound of $\beta_0/2M^2 \leq 0.01064$, where $M$ is the black hole mass. M87 has a mass of $6.5 \times 10^9 M_{\odot}$, hence producing an upper bound of $\beta_0 \sim 10^{78}$ \cite{tamburini2022constraining}. 

		\subsection{Modification to phase space}

Chang et al. in Ref. \refcite{chang2002effect} reports that the RHS of \eqref{eq:delXdelP_quadratic GUP} implies a $p$-dependence in the ``effective value of $\hbar$". This in turn implies that the size of the unit cell that each quantum state occupies in phase space is also $p$-dependent. The modified phase space volume thus takes the form
\begin{equation}
\frac{d^{D} \mathbf{x} d^{D} \mathbf{p}}{\left(1+\beta p^{2}\right)^{D}} \rightarrow \frac{V}{(2 \pi \hbar)^{D}} \frac{d^{D} \mathbf{p}}{\left(1+\beta p^{2}\right)^{D}}  \label{eq:final_qgup_phase_space}
\end{equation}
where the volume $V$ arises from the trivial configuration space integration $\int d^{D} x =V$, and $(2 \pi \hbar)^{-D}$ arises when considering quantum mechanical systems. The phase space volume above has been checked by Chang et al. to contain an unchanging number of states as the volume evolves in time (an analog of the Liouville theorem). 
	
		\subsection{Modified equation of state}

Because no white dwarfs have been observed to surpass the Chandrasekhar mass limit \cite{mathew2014general, kepler2007white, bedard2017measurements}, the Chandrasekhar EoS is often used to describe the relation between the pressure and energy density of white dwarf matter. Said matter is assumed to consist of a highly-degenerate Fermi gas of non-interacting electrons \cite{fantoni2017white, hamada1961models}, exerting an outward pressure that counteracts the object's self-gravity \cite{jackson2005compact}. For a purely Newtonian investigation of white dwarf structure, it is sufficient to assume that only the nucleons contribute to the energy density, but our general relativistic treatment requires we consider contributions from both nucleons and electrons \cite{carvalho2018general}.

The modified phase space volume affects the statistical mechanics of Fermi particles, which by extension, affects the thermodynamic properties involved in constructing the EoS. To derive the modified properties, we begin with the grand canonical partition function \cite{sagert, jackson2005compact, balian1999stars} for a a system of particles with microstates $j$, energies $E_j$, and number of particles $n_j$, given as
\begin{equation}
Z=\sum_{n_{j}} \prod_{j}\left[e^{\left(\mu-E_{j}\right) / k_{B} T}\right]^{n_{j}} \rightarrow \ln Z=\sum_{j} \ln \left[1+e^{\left(\mu-E_{j}\right) / k_{B} T}\right]
\end{equation}
where $Z$ is summed over all possible microstates. $\mu$ is the chemical potential, $k_{B}$ is Boltzmann’s constant, and T is the temperature of the system. On the RHS, we have expressed $Z$ in terms of its natural logarithm, from which we can derive the thermodynamic properties \cite{bertolami2010towards}. For systems of large volume, we rewrite the summation as an integral over the modified phase space \eqref{eq:final_qgup_phase_space}, such that $\ln Z$ becomes
\begin{equation}
\ln Z= \frac{gV}{(2 \pi \hbar)^{3}}\int_{0}^{\infty} \frac{4 \pi p^{2} d p}{\left(1+\beta p^{2}\right)^{3}} \ln \left[1+e^{\left(\mu-E\right) / k_{B} T}\right] \label{eq:lnZ}
\end{equation}
where we have used $D = 3$, and $\int d^{3} p$  has been reduced to $4 \pi \int p^{2} d p$ assuming spherical symmetry. $g$ is the degeneracy factor, equal to 2 for electrons \cite{sagert}. $E$ is given by the energy-momentum dispersion relation of arbitrary relativity,
\begin{equation}
E=\sqrt{p^{2} c^{2}+m_{e}^{2} c^{4}}
\end{equation}
where $m_e$ is the electron mass. Following methods discussed in Ref. \citen{moussa}, we can derive from \eqref{eq:lnZ} the particle number density $n$, electron energy density $\epsilon$, and pressure $P$:
\begin{align}
n = \frac{1}{\pi^2 \hbar^3} &\int_{0}^{\infty} \frac{1}{1+e^{(E-\mu) / k_{B} T}} \frac{p^{2} d p}{\left(1+\beta p^{2}\right)^{3}} \label{eq:number_density} \\
\epsilon= \frac{1}{\pi^2 \hbar^3} &\int_{0}^{\infty} \frac{E}{1+e^{(E-\mu) / k_{B} T}} \frac{p^{2} d p}{\left(1+\beta p^{2}\right)^{3}} \label{eq:energy_density}\\
P = \frac{k_{B} T}{\pi^2 \hbar^3} &\int_{0}^{\infty} \ln \left[1+e^{(\mu-E) / k_{B} T}\right] \frac{p^{2} d p}{\left(1+\beta p^{2}\right)^{3}} \label{eq:pressure}
\end{align}
In \eqref{eq:number_density} and \eqref{eq:energy_density}, we encounter the Fermi-Dirac distribution $f(E)$, 
\begin{equation}
f(E) = (1+e^{(E-\mu) / k_{B} T})^{-1}
\end{equation}
$f(E)$ can be found in the pressure expression by integrating \eqref{eq:pressure} by parts, which yields:
\begin{equation}
P = \frac{1}{\pi^2 \hbar^3} \int_{0}^{\infty} \left[\frac{\arctan(\sqrt{\beta}p)}{8\beta^{3/2}} + \frac{p(\beta p^2 -1)}{8\beta (1 + \beta p^2)^2}\right] \left\lbrace \left[\frac{1}{1+e^{(E-\mu) / k_{B} T}}\right]\frac{c^2 p}{E}dp \right\rbrace \label{eq:pressure_byparts}
\end{equation}
From \eqref{eq:number_density}, we can construct an expression for the mass density $\rho$:
\begin{equation}
\rho = \mu_e m_u n
\end{equation}
where $\mu_e$ is the molecular weight per electron, and $m_u$ is the atomic mass unit. For white dwarfs made of He, C, and O, $\mu_e \approx 2$ \cite{boshkayev}, which is the specific $\mu_e$ we will use for the rest of the paper. Finally, the total energy density $\epsilon_t$ is defined as
\begin{equation}
\epsilon_t = c^2 \rho + \epsilon \label{eq:eps_t} \\
\end{equation}
where the first term on the right-hand side corresponds to the nucleon contribution, while the second term is the electron contribution.

The Chandrasekhar EoS is therefore obtained by calculating \eqref{eq:eps_t} and \eqref{eq:pressure_byparts} at various $\mu$, and subsequently interpolating those values to obtain a function of the form $\epsilon_t (P)$. For convenience in numerical calculations, we further obtain the dimensionless EoS by using the quantities:
\begin{equation}
\tilde{\epsilon}=\frac{\epsilon_{t}}{\epsilon_{0}} ; \quad \tilde{P}=\frac{P}{\epsilon_{0}} ; \quad \epsilon_{0}=\frac{m_{e}^{4} c^{5}}{\pi^{2} \hbar^{3}} \label{eq:dimensionless_eos_relations}
\end{equation}
where $\epsilon_0$ has units of pressure.
	
	\section{Approaches to calculate the equation of state}
	\label{section: section_3}	
	
	\subsection{Perturbative approach}
		
The pressure integral \eqref{eq:pressure_byparts} is arguably a complicated integral to solve numerically, hence the need arises to rewrite the integral toward a more suitable form for numerical integration. A first, reasonable approach would be to treat $\beta$ as a perturbative parameter, thus allowing us to expand the thermodynamic integrals around $\beta$.

%\subsubsection{The pressure integral}	
We first notice that the GUP modification to \eqref{eq:pressure_byparts} is found in the factor in square brackets before $f(E)$. Its corresponding Taylor series is given as
\begin{equation}
\frac{\arctan(\sqrt{\beta}p)}{8\beta^{3/2}} + \frac{p(\beta p^2 -1)}{8\beta (1 + \beta p^2)^2} \approx \frac{1}{3}p^3 - \frac{3}{5}\beta p^5 +\frac{6}{7}\beta^2 p^7 - \frac{10}{9}\beta^3 p^9 + \frac{15}{11}\beta^4 p^{11} + ... \label{eq:v_taylor}
\end{equation}
which is similar in form to the series obtained by Wang et al. in Ref. \citen{wang}. Moussa in Ref. \citen{moussa} only kept terms up to $\mathcal{O}(\beta)$ in the modified zero-temperature EoS; in this paper, we will use higher-order terms.

Plugging the Taylor series back into \eqref{eq:pressure_byparts}, we obtain:
\begin{align}
\begin{split}
P &=\frac{1}{\pi^2 \hbar^3} \int_{0}^{\infty} \left[  \frac{1}{3}p^3 - \frac{3}{5}\beta p^5 +\frac{6}{7}\beta^2 p^7 - \frac{10}{9}\beta^3 p^9 + \frac{15}{11}\beta^4 p^{11} + ...              \right] \\
&\qquad \times \left\lbrace \left[\frac{1}{1+e^{(E-\mu) / k_{B} T}}\right]\frac{c^2 p}{E}dp \right\rbrace
\end{split} \label{eq:taylor_p} \\
\begin{split}
  &= P_0 + P_1 + P_2 + P_3 + P_4 + \hdots = \sum_{j = 0}^{\infty} P_j
\end{split}
\end{align}
We are interested in deriving an expression for $P_j$ to be able to approximate $P$ up to any order $j$. We do so by introducing the following substitutions \cite{faussurier2016equation}
\begin{align}
\begin{split}
E = m_e c^2 (\theta x + 1); \quad E_k = E - m_e c^2; \quad \mu' = \mu - m_e c^2 \\
\theta = \frac{k_B T}{m_e c^2}; \quad x = \frac{E_k}{k_B T}; \quad \eta = \frac{\mu'}{k_B T} \\
p = \sqrt{2m_ek_B T x}\left(1+ \theta x/2\right)^{1/2}; \quad dp = k_B T \frac{E}{c^2 p} dx
\end{split} \label{eq:gup_integral_subs}
\end{align}
such that $P_j$ is written as
\begin{equation}
P_j = C_j \frac{m_e^4 c^5}{\pi^2 \hbar^3} 2^{\left[\frac{3}{2} + j \right]} \theta^{\left[\frac{5}{2} + j \right]} \tilde{\beta}^{2j} \sum_{i=1}^{j+2} D_i F_{\left(\frac{1}{2} + j + i\right)} \label{eq:p_formula}
\end{equation}
where $C_j$ is the $j^{th}$ coefficient in the Taylor series \eqref{eq:v_taylor}, $D_i$ is the $i^{th}$ term in the polynomial $\left(1+\frac{\theta}{2}\right)^{j+1}$, $\tilde{\beta}$ is related to $\beta$ \& $\beta_0$ by
\begin{equation}
\tilde{\beta} = m_e c\sqrt{\beta} = \frac{m_e l_p c }{\hbar}\sqrt{\beta_0}
\end{equation}
and
\begin{equation}
F_k = \int_{0}^{\infty} \frac{x^k \sqrt{1+\theta x/2}}{1 + e^{x-\eta}} dx \label{eq:gen_fermi_dirac}
\end{equation}
is known as the Generalized Fermi-Dirac integral \cite{faussurier2016equation, boshkayev}. The presence of $F_k$ in the above expressions is particularly convenient, as numerous methods already exist to calculate $F_k$ for any value of $k$ and $\eta$, such examples being References \citen{fukushima2015precise1,fukushima2015precise2,fukushima2015precise3}.

We note however, that using the lowest-ordered terms in the expansion is only valid for $\tilde{\beta} << 1$. The approximation reasonably breaks down for $\tilde{\beta} \gtrsim 1$, because the coefficient $ \tilde{\beta}^{2j}$ in \eqref{eq:p_formula} would create higher-order terms that are larger in magnitude compared to preceding terms, i.e. $P_{j-1} < P_j$.

For the number and energy density integrals \eqref{eq:number_density} \& \eqref{eq:energy_density}, we can expand the factor $p^2 (1+\beta p^2)^{-3}$ through the following Taylor series
\begin{equation}
\frac{p^2}{(1+\beta p^2)^3} \approx p^2 -3\beta p^4 + 6\beta^2p^6 - 10\beta^3 p^8 + 15\beta^4 p^{10} + ... \label{eq:taylor_n}
\end{equation}
and thus derive the following expressions for $n_j$ and $\epsilon_j$:
\begin{equation}
\begin{split}
n_j &= G_j \frac{m_e^3 c^3}{\pi^2 \hbar^3} 2^{\left[\frac{1}{2} + j \right]} \theta^{\left[\frac{3}{2} + j \right]} \tilde{\beta}^{2j} \sum_{i=1}^{j+2} H_i F_{\left(j + i - \frac{1}{2} \right)} \\
\epsilon_{j} &= G_j \frac{m_e^4 c^5}{\pi^2 \hbar^3} 2^{\left[\frac{1}{2} + j \right]} \theta^{\left[\frac{3}{2} + j \right]} \tilde{\beta}^{2j} \sum_{i=1}^{j+3} J_i F_{\left(j + i - \frac{1}{2} \right)}
\end{split}
\end{equation}
where $G_j$ is the $j^{th}$ coefficient in the Taylor series \eqref{eq:taylor_n}, $H_i$ is the $i^{th}$ term in the polynomial $(1+\theta)\left(1+\frac{\theta}{2}\right)^{j}$, and $J_i$ is the $i^{th}$ term in the polynomial $(1+\theta)^2\left(1+\frac{\theta}{2}\right)^{j}$.

\subsection{Visualizing the perturbative equation of state}

\begin{figure}[t]
		\centering
		\includegraphics[width=1\textwidth]{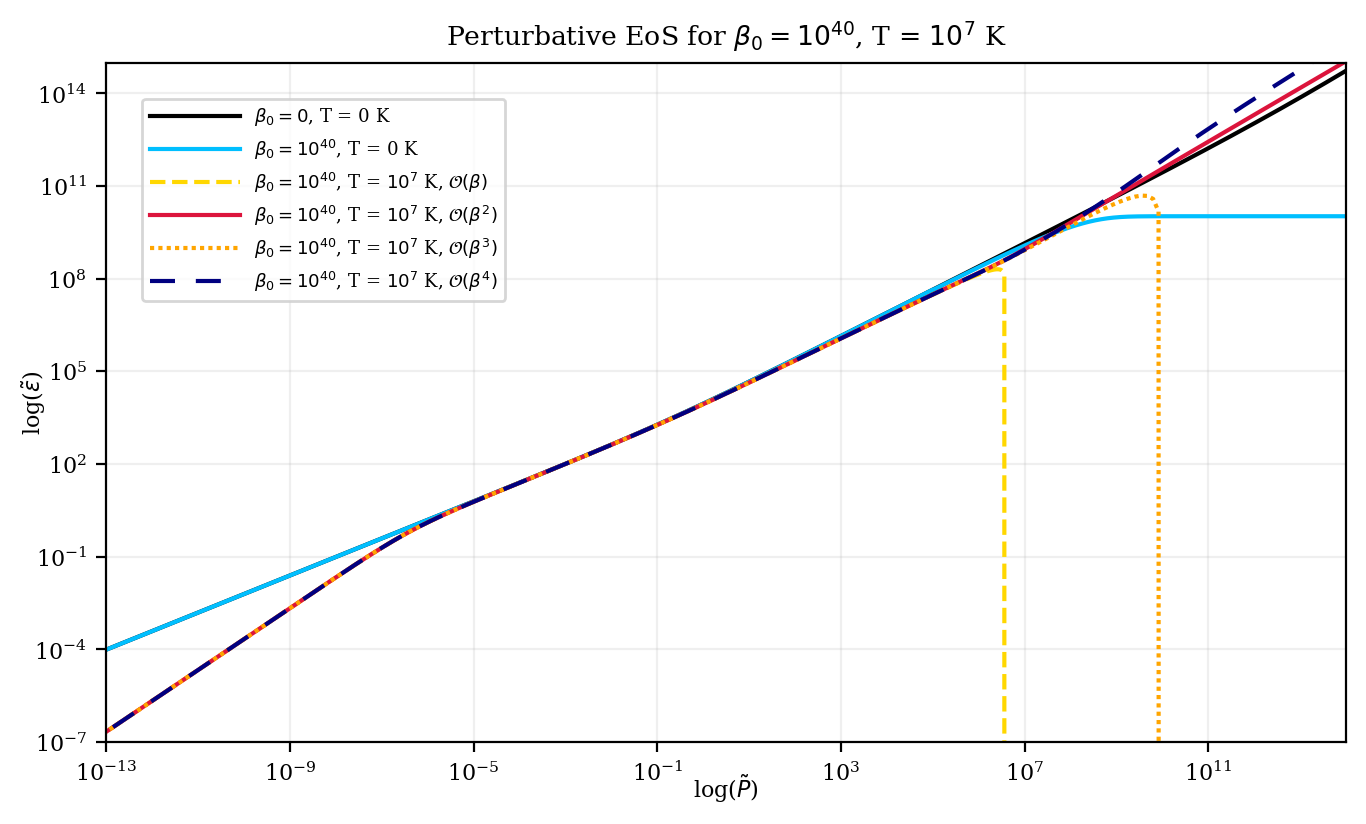}
		\caption{Perturbative EoS at various orders of approximation, for $\beta_0  = 10^{40}$ and T = $10^7$ K.}
		\label{fig:perturbative_eos}
	\end{figure}

Using \eqref{eq:dimensionless_eos_relations}, we can calculate the EoS at various orders of approximation. The resulting graphs are found on Figure \ref{fig:perturbative_eos}, where we have used $\beta_0 = 10^{40}$ (or $\tilde{\beta} \sim 10^{-3}$) and $T = 10^7$ K. Due to limitations in computation, we limit the approximation to fourth order. For comparison purposes, we superimpose the EoS for $\beta_0 = 10^{40}$ \& $T = 0$ K (solid sky-blue line), which starts as coinciding with the regular zero temperature EoS ($\beta_0 = 0$ \& $T = 0$ K) as $\tilde{P}$ increases, but later saturating\footnote{``saturate" as used in the context of the EoS is a term that is lifted from the GUP discussions of Ref. \citen{mathew}.} toward a constant energy density, thus indicating a stiffer EoS \cite{niemeyer2002varying}. Here, ``stiffen" refers to the continuously large increases in pressure corresponding to minimal changes in energy density, indicative of a gas that is harder to compress \cite{rezzolla2013relativistic}.

The perturbative EoS (for all orders of approximation) exhibits the expected thermal deviation at low pressures \cite{boshkayev}, diverging from the ideal case with a steeper slope (indicating a softer EoS), before converging with the ideal case at $\tilde{\epsilon} \gtrsim 10^{-1}$. In the high pressure regime, we encounter the following behaviors: The $\mathcal{O}(\beta)$ EoS (dashed yellow line) dips and shoots downward at $\tilde{\epsilon} \sim 10^{8}$ because EoS values beyond this point are negative -- thus becoming undefined in the logarithmic plot. This may be attributed to the negative sign of $\beta^1$ terms in the Taylor series \eqref{eq:taylor_p} and \eqref{eq:taylor_n}. Through the plot, we can infer that terms proportional to $\beta$ are larger in magnitude compared to terms proportional to $\beta^0$, with the former dominating the latter where the EoS starts to dip. For the $\mathcal{O}(\beta^2)$ EoS however (solid red line), values remain positive all throughout, gaining a larger slope than the ideal EoS at $\tilde{\epsilon} \gtrsim 10^{11}$, once again indicative of the dominance of the last truncated term (which now has a positive sign). Furthermore, the occurence of a larger slope is reminiscent of behavior found in the zero temperature EoS with \textit{linear} GUP, as explored by Ref. \citen{abac, abac2}. These behaviors are repeated (and more pronounced) for  $\mathcal{O}(\beta^3)$ and $\mathcal{O}(\beta^4)$, each dictated by the sign of the largest-ordered term.

One can observe that the high-pressure deviations occur in the region where saturation also begins for the zero temperature EoS with GUP. This allows us to further infer that as the order increases, and thermodynamic integrals ($n_j, \epsilon_j, P_j$) of higher order $j$ are taken into account, the perturbative EoS should approach the zero temperature GUP case. We note however, that further increasing the order $j$ would entail a Fermi-Dirac integral of the form
\begin{equation}
\int_0^{\infty} x^{j \rightarrow \infty} \bigg[\text{other terms}\bigg] dx
\end{equation}
which is quite an impractical integral to solve. We are also hampered by the conflicting behaviors shown by the EoS at various orders, due to the alternating signs of the Taylor series expansion. Choosing a large $j$ does not eliminate the fact that the EoS behaves differently for $j+1$. We are therefore urged to forego the approximation in the interest of a full numerical approach to solve the thermodynamic integrals.

\subsection{Non-perturbative approach}
		
To remedy the above problems, here we calculate the modified EoS in its exact form. We begin with the modified number and energy density integrals \eqref{eq:number_density} \& \eqref{eq:energy_density}, and the pressure integrated by parts \eqref{eq:pressure_byparts}. Using \eqref{eq:gup_integral_subs}, while choosing the equivalent $p$ substitution $p = m_e c\sqrt{(\theta x + 1)^2 -1}$, we obtain the integrals:
\begin{align}
\begin{split}
n &= \frac{m_e^3 c^3}{\pi^2\hbar^3}\theta \int_{0}^{\infty}  \left[ \frac{(\theta x + 1)}{1 + e^{x-\eta}}\right]  \frac{\sqrt{(\theta x + 1)^2 -1} }{\{1 + \tilde{\beta}^2\left[(\theta x + 1)^2 -1\right]\}^3} \ dx
\end{split} 
\\[2ex]
\begin{split}
\epsilon &= \frac{m_e^4 c^5}{\pi^2\hbar^3}\theta \int_{0}^{\infty}  \left[ \frac{(\theta x + 1)^2}{1 + e^{x-\eta}}\right]  \frac{\sqrt{(\theta x + 1)^2 -1} }{\{1 + \tilde{\beta}^2\left[(\theta x + 1)^2 -1\right]\}^3} \ dx
\end{split} 
\\[2ex]
\begin{split}
P &= \frac{m_e^4 c^5}{\pi^2\hbar^3} \frac{\theta}{8\tilde{\beta}^3} \int_{0}^{\infty} \Bigg\{ \left[ \frac{1}{1 + e^{x-\eta}}\right] \arctan \left[ \tilde{\beta}\sqrt{(\theta x + 1)^2 -1} \right] \\
		&\qquad + \left[ \frac{\tilde{\beta}^2 \left[(\theta x + 1)^2 -1\right] - 1}{1 + e^{x-\eta}}\right] \frac{\tilde{\beta} \sqrt{(\theta x + 1)^2 -1}}{\{1 + \tilde{\beta}^2\left[(\theta x + 1)^2 -1\right]\}^2} \Bigg\} \ dx 
\end{split}
\end{align}
Although this exact approach does not share the convenience of the perturbative EoS and the pre-existing algorithms to solve $F_k$, the numerical work required is somewhat reduced through our efforts to re-express the integrals in terms of kinetic energy $x$, instead of the momentum $p$, the latter being inconveniently nested in $E$ and $f(E)$ throughout the original integrals. More importantly, this approach should be applicable for arbitrary temperature $T$ and GUP parameter $\tilde{\beta}$.

\begin{figure}[t]
    \centering
    \includegraphics[width=1\textwidth]{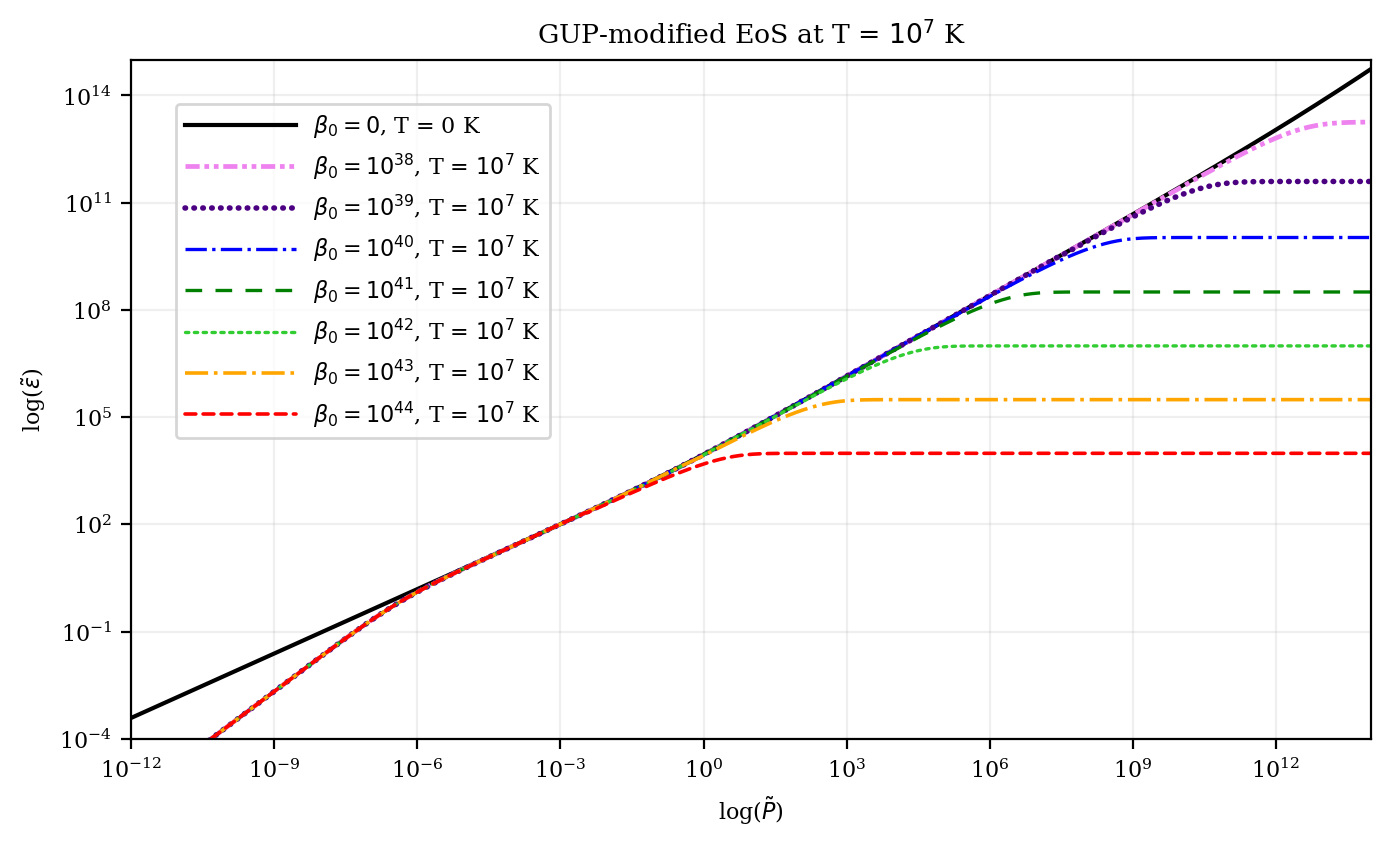}
    \caption{The GUP-modified finite temperature EoS at T = $10^7$ K.}
    \label{fig:EoS_QGUP_T}
\end{figure}
 
Using the above integrals to calculate the EoS allows us to obtain the plot in Figure \ref{fig:EoS_QGUP_T}, where we have used the GUP parameters in the range $10^{38} \leq \beta_0 \leq 10^{44}$ (as employed by Mathew and Nandy in Ref. \citen{mathew}, all of which falling within the estimated boundaries in Section \ref{subsection:quadratic_GUP}), with temperature $T = 10^7$ K. In the low-pressure regime, we find the expected thermal deviation of a softer EoS, occuring for all $\beta_0$. In the high-pressure regime, the EoS stiffens and saturates toward constant energy densities at high pressures. The ``saturation point," i.e. the point at which saturation toward constant energy densities starts to occur, is dependent on the magnitude of $\beta_0$. Larger $\beta_0$ leads to stronger deviations, resulting in a saturation point that occurs at smaller pressure. Conversely, smaller $\beta_0$ leads to a higher saturation point. We can also confirm the validity of our results in Figure \ref{fig:EoS_QGUP_T} by comparing them with the GUP EoS plots at zero temperature as produced by Mathew and Nandy in Ref. \citen{mathew}. Sans the thermal deviation in the former figure, both exhibit identical GUP effects.
\\

	\section{Quadratic GUP and white dwarf structure}
	\label{section: section_4}
	
		\subsection{The Tolman-Oppenheimer-Volkoff equations}
		
		The Tolman-Oppenheimer-Volkoff (TOV) equations are a system of first-order ordinary differential equations (ODEs) describing the pressure $P$ and mass $M$ of a spherically-symmetric, general relativistic star from center to surface \cite{mathew2014general}. To derive the ODEs, Tolman \cite{tolman1939static}, Oppenheimer and Volkoff \cite{oppenheimer1939massive} solved the Einstein field equations for a static and isotropic star made of a perfect fluid, whose exterior spacetime is described by the Schwarzschild metric \cite{glendenning}. The ODEs read as
\begin{align}
\frac{d P}{d r} &=-\frac{G M \epsilon_t(r)}{c^{2} r^{2}}\left[1+\frac{P}{\epsilon_t(r)}\right]\left[1+\frac{4 \pi r^{3} P}{M c^{2}}\right]\left[1-\frac{2 G M}{c^{2} r}\right]^{-1} \label{eq:dp/dr_GR}\\
\frac{d M}{d r} &=\frac{4 \pi r^{2} \epsilon_t(r)}{c^{2}} \label{eq:dm/dr}
\end{align}
where $\epsilon_t(r)$ is the interpolated EoS, a function of the pressure $P$. \eqref{eq:dp/dr_GR} is a statement of hydrostatic equilibrium, i.e. the balance between a star's pressure and self-gravity. It is a monotonically decreasing function, vanishing at the stellar radius $R_{\star}$. \eqref{eq:dm/dr} is the mass continuity equation, which tells us that a star's mass increases with its radius, capping off at the stellar radius with a total mass $M_{\star}$. As in previous studies involving the stellar structure equations \cite{abac, abac2021effects, abac2023stability, abac2021implications}, these ODEs are to be solved numerically, since there are no analytic solutions \cite{silbar2004neutron}; for this paper we use the standard fourth-order Runge-Kutta method (RK4) as employed by Refs. \citen{silbar2004neutron, carvalho2018general, sagert,jackson2005compact, abac}. To do so, we employ the boundary conditions $P(r=0)=P_0$, where $P_0$ is the central pressure, and $M(r=0)=0$.

The factors in square brackets on \eqref{eq:dp/dr_GR} can be considered positive GR corrections to the stellar pressure \cite{carvalho2018general}. Without these corrections, we recover the Newtonian structure equations, given by the pressure equation 
\begin{equation}
\frac{d P}{d r} =-\frac{G M \epsilon_t(r)}{c^{2} r^{2}} \label{eq:dp/dr_NC}
\end{equation}
and the same mass equation in \eqref{eq:dm/dr}. Comparing \eqref{eq:dp/dr_GR} and \eqref{eq:dp/dr_NC}, one can conclude that the gravitational pull is stronger in GR \cite{silbar2004neutron}. This Newtonian approximation to the TOV equations is sufficient to study the essential features of white dwarfs \cite{boshkayev}, since most are found within the $0.5$ to $0.7$ solar mass range \cite{kepler2007white}. However, the TOV equations in their exact form are needed in studying the most massive white dwarfs, as these objects significantly warp spacetime \cite{glendenning, jackson2005compact, silbar2004neutron}.

To avoid dealing with astronomically-large quantities of mass and radius, which may incur large and unnecessary numerical errors \cite{jackson2005compact}, we rewrite the TOV equations into their computationally-efficient dimensionless forms:
\begin{align}
\frac{d \tilde{P}}{d \tilde{r}} &=-\frac{\tilde{M}(\tilde{r}) \tilde{\epsilon}(\tilde{r})}{\tilde{r}^{2}}    \left[1+\frac{\tilde{P}(\tilde{r})}{\tilde{\epsilon}(\tilde{r})}\right]\left[1+\frac{\alpha \tilde{r}^{3} \tilde{P}(\tilde{r})}{R_{0}^{3}\tilde{M}(\tilde{r})} \right] \left[1-\frac{2 \tilde{M}(\tilde{r})}{\tilde{r}}\right]^{-1} \label{eq:dp/dr_GR_dimensionless} \\
\frac{d \tilde{M}}{d \tilde{r}} &=\frac{\alpha}{R_{0}^{3}} \tilde{r}^{2} \tilde{\epsilon}(\tilde{r}) \label{eq:dm/dr_GR_dimensionless}
\end{align}
where
\begin{equation}
\alpha=\frac{4 \pi \epsilon_{0}}{M_{\odot} c^{2}}, \quad \tilde{r}=\frac{r}{R_{0}}, \quad R_{0}=\frac{c^{2}}{G M_{\odot}}, \quad \tilde{M}=\frac{M}{M_{\odot}}
\end{equation}
and $ \tilde{\epsilon}(\tilde{r})$ is given by \eqref{eq:dimensionless_eos_relations}. It should be easy to see that the Newtonian approximation to \eqref{eq:dp/dr_GR_dimensionless} takes the form
\begin{equation}
\frac{d \tilde{P}}{d \tilde{r}} =-\frac{\tilde{M}(\tilde{r}) \tilde{\epsilon}(\tilde{r})}{\tilde{r}^{2}} \label{eq:dp/dr_NC_dimensionless}
\end{equation}
while the Newtonian mass retains the form of \eqref{eq:dm/dr_GR_dimensionless}.

		\subsection{Newtonian solutions}

\begin{figure}[t!]
    \centering
    \includegraphics[width=1\textwidth]{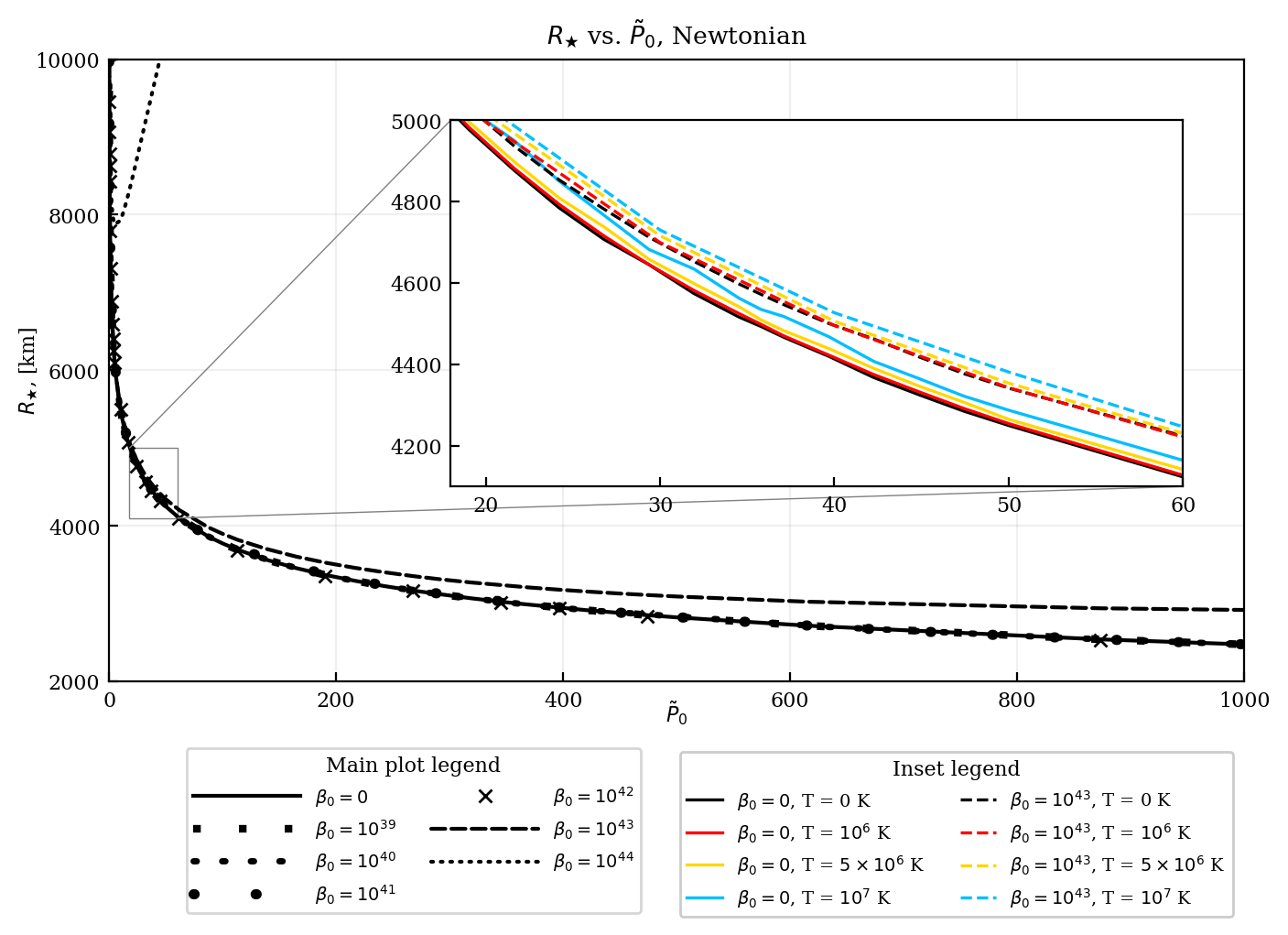}
    \caption{$R_{\star}$ vs. $\tilde{P}_0$ for GUP-modified Newtonian white dwarfs at finite temperatures.}
    \label{fig:NC_rvP_QGUP}
\end{figure}

We first solve the Newtonian ODEs for a large number of central pressures to obtain a variety of stellar masses and radii. A stellar radius versus central pressure plot is shown on Figure \ref{fig:NC_rvP_QGUP}, where we see that similar radii are produced for $10^{39} \leq \beta_0 \leq 10^{42}$, decreasing as $\tilde{P}_0$ increases. The same behavior is observed for $\beta_0 = 10^{43}$, albeit with slightly larger radii obtained across central pressures. For $\beta_0 = 10^{44}$ however, the radii rapidly increase even at small central pressures. It is also at small central pressures that we see the effects of temperature. As shown on the inset, higher temperatures lead to slightly larger radii, as is observed even for regular white dwarfs \cite{boshkayev}. At larger pressures however, this thermal effect disappears.

\begin{figure}[t!]
    \centering
    \includegraphics[width=1\textwidth]{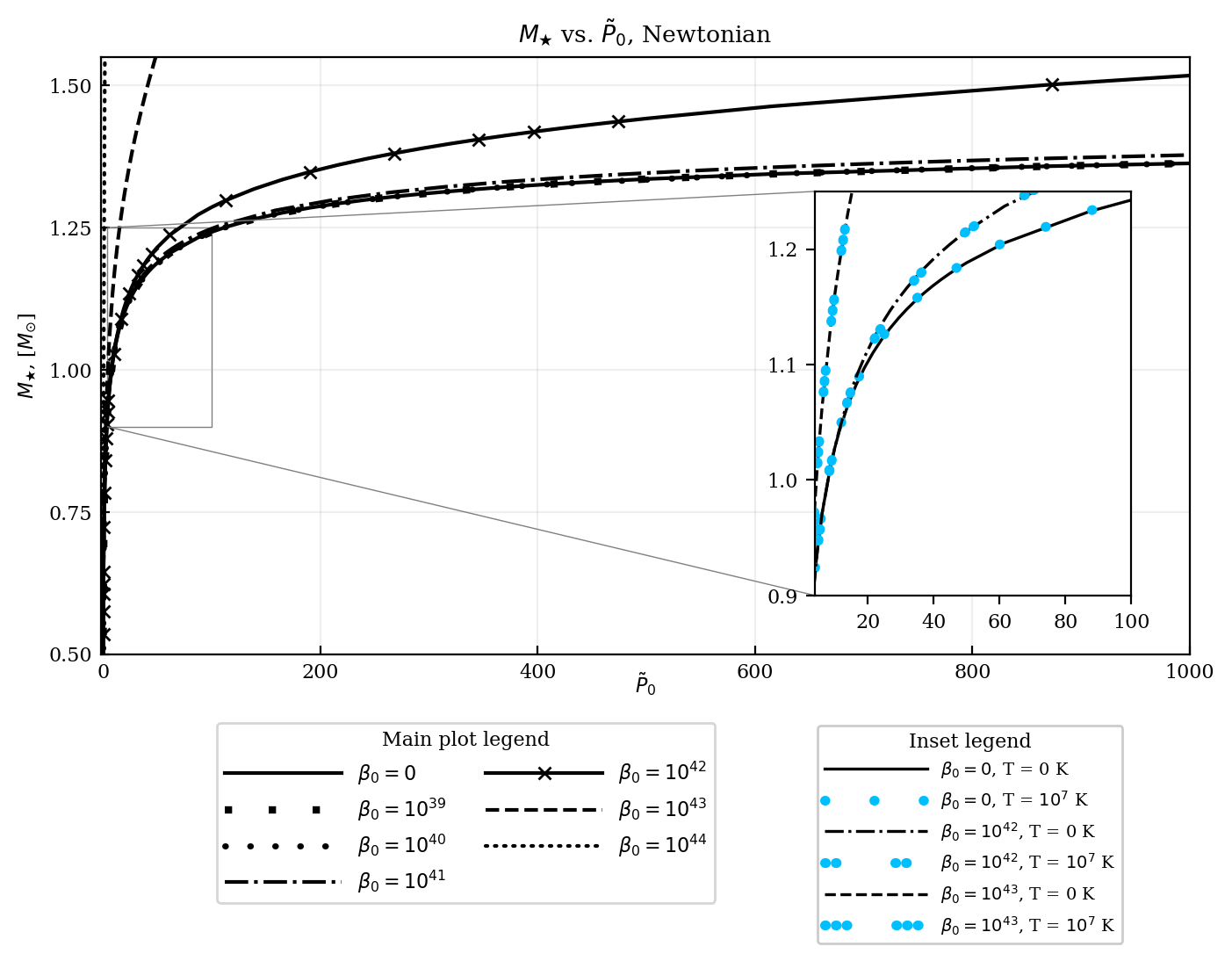}
    \caption{$M_{\star}$ vs. $\tilde{P}_0$ for GUP-modified Newtonian white dwarfs at finite temperature.}
    \label{fig:NC_MvP_QGUP}
\end{figure}

A stellar mass versus central pressure plot is also shown on Figure \ref{fig:NC_MvP_QGUP}, where masses produced for $\beta_0 = 10^{39}, 10^{40}$ are similar to those found in the ideal case, both of which plateauing at the Chandrasekhar mass limit of $M_{\text{Ch}} = 1.456M_{\odot}$  \cite{chandrasekhar, sagert, jackson2005compact, balian1999stars, silbar2004neutron} as $\tilde{P}_0$ increases. Masses for $\beta_0 = 10^{41}$ are slightly elevated from the previous $\beta_0$ values, while masses for $\beta_0 = 10^{42}$ forego the plateau and increase steadily instead. This effect is also observed for larger $\beta_0$ but stronger, with masses rapidly increasing from the outset -- a strange result as opposed to the constancy of the Chandrasekhar mass in the limit of infinite central pressure (at least within the Newtonian context). As shown on the inset, no thermal effect can be seen for the masses across all $\beta_0$.

\begin{figure}[t!]
    \centering
    \includegraphics[width=1\textwidth]{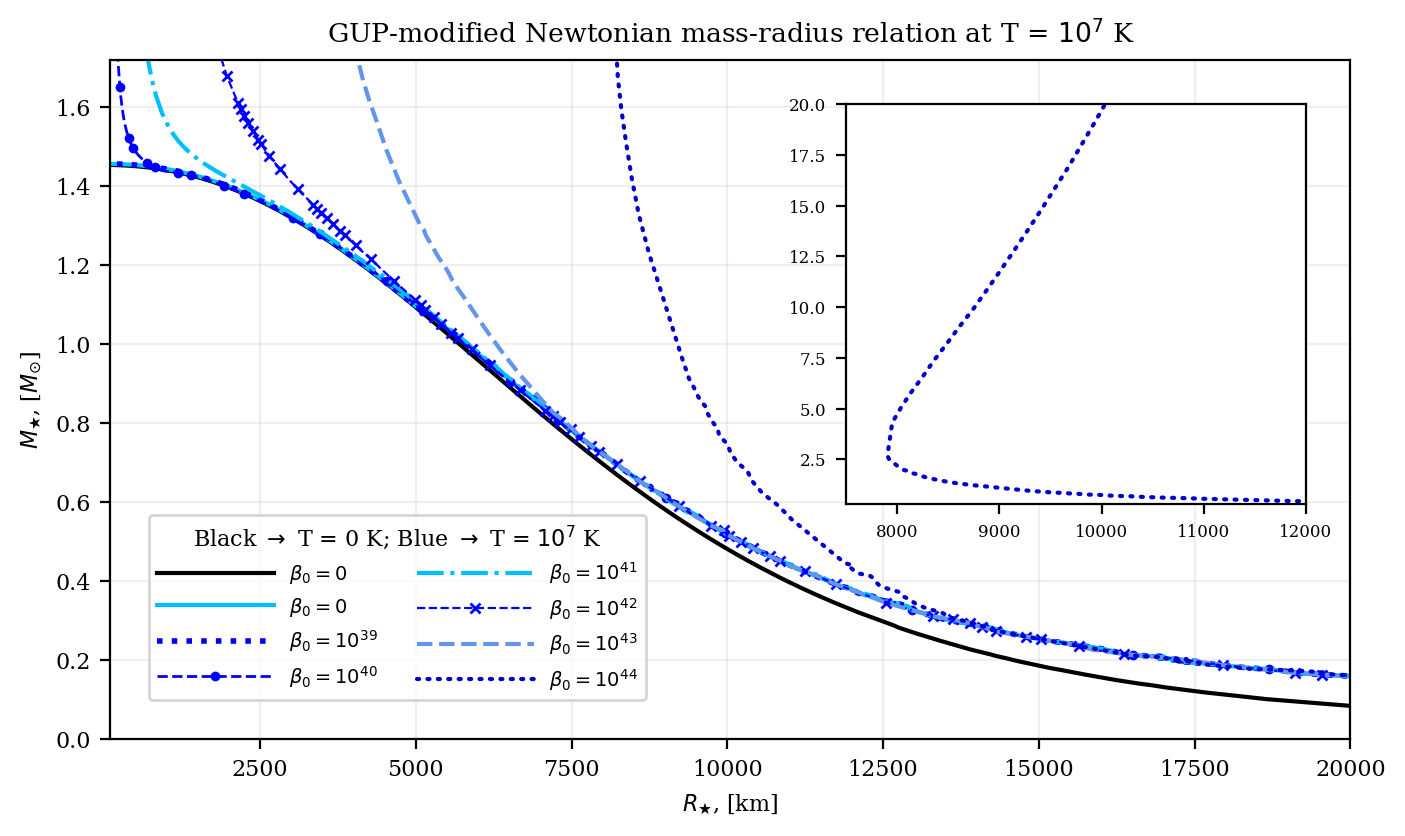}
    \caption{GUP-modified mass-radius relations for Newtonian white dwarfs at T = $10^7$ K.}
    \label{fig:mass-radius_qgup_t}
\end{figure}

\begin{figure}[t!]
    \centering
    \includegraphics[width=1\textwidth]{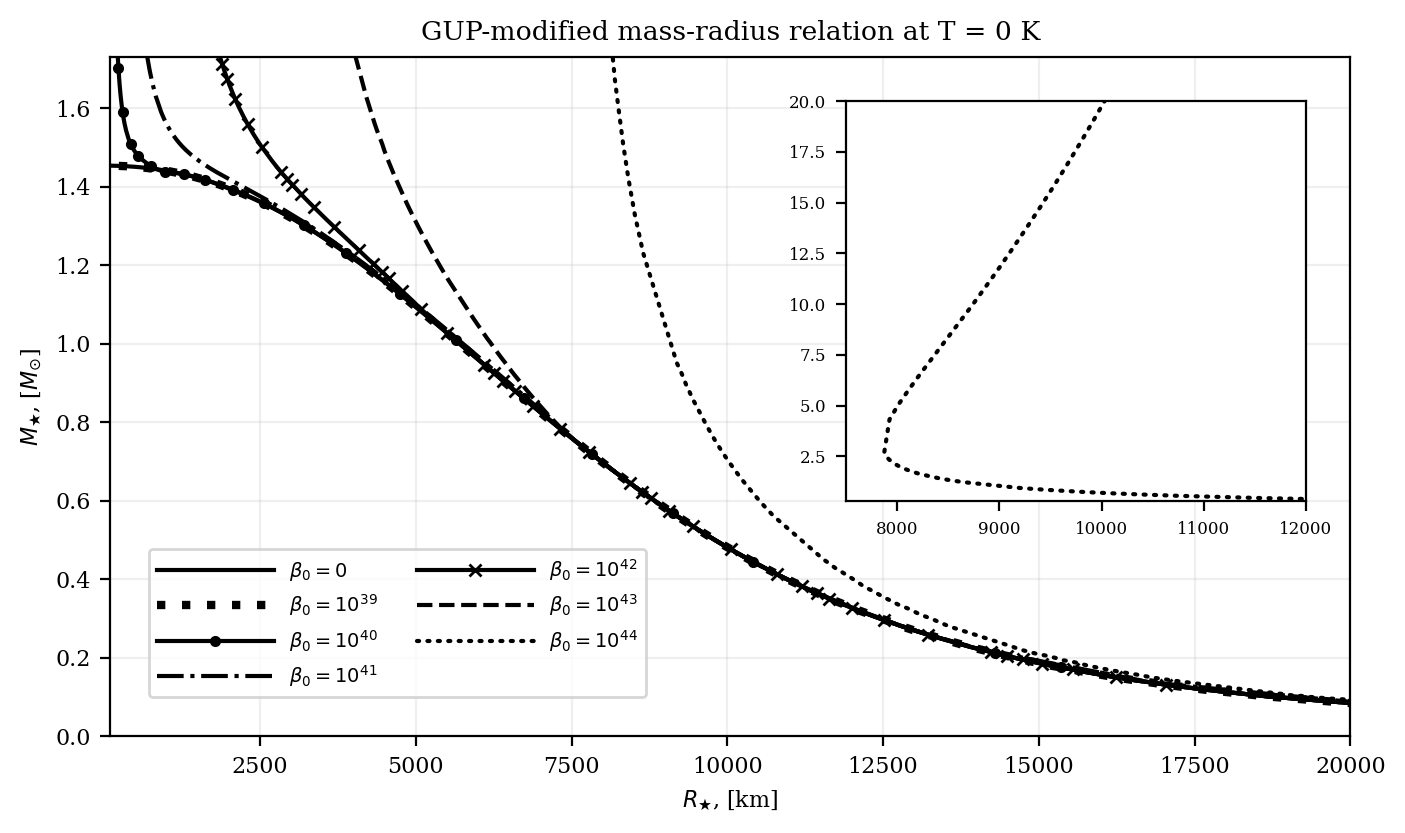}
    \caption{GUP-modified mass-radius relations for Newtonian white dwarfs at zero temperature.}
    \label{fig:mass-radius_qgup_zero}
\end{figure}

Finally, by plotting the stellar masses against the radii, we obtain the modified mass-radius relations in color on Figure \ref{fig:mass-radius_qgup_t}, superimposed over the ideal case in black. The relations obtained using our modified EoS at various $\beta_0$ all exhibit the expected thermal deviation in the low-mass regime, where masses found to the right of the plot are slightly larger than the ideal case \cite{boshkayev}. In the high mass regime, the relation for $\beta_0 = 10^{39}$ closely follows the ideal case, while relations of larger $\beta_0$ deviate and extend toward infinite mass values. Larger $\beta_0$ corresponds to stronger deviations, in the sense that more low-mass white dwarfs stray from the ideal behavior, as the relation is lifted from the ideal case. On the inset is the relation for $\beta_0 = 10^{44}$ with the y-axis limits extended toward large masses beyond the Chandrasekhar limit, showing that indeed, both masses and radii increase without bounds. The relations produced here resemble those produced in Ref. \citen{mathew} (see Figure \ref{fig:mass-radius_qgup_zero}), only differing in the thermal deviation found in the low-mass regime. They also confirm the results found from the heuristic calculations of Ref. \citen{ong}, of white dwarfs being able to ``bounce" from gravitational collapse, this bouncing effect being the tendency of the modified relations to approach a limiting mass toward the left of the graph, only to make a turn for the right as the central pressure approaches infinity.

The absence of a mass limit for white dwarfs is, as far as current observations are concerned \cite{kilic2007lowest, kepler2007white, bedard2017measurements}, a nonphysical result. Beyond the mass limit, white dwarfs are expected to collapse into neutron stars and black holes \cite{silbar2004neutron, vidana2018short, shapiro2008black, glendenning, hartle}. Additionally, as pointed out by References \citen{moussa} and \citen{ong}, the GUP phenomenon of allowing arbitrarily large values of mass and radius is also far removed from the reality that white dwarfs from observation have smaller radii than what is predicted by theory \cite{ mathews2006analysis, camacho2006white, provencal2002procyon}.

\subsection{General relativistic solutions}

\begin{figure}[t!]
    \centering
    \includegraphics[width=1\textwidth]{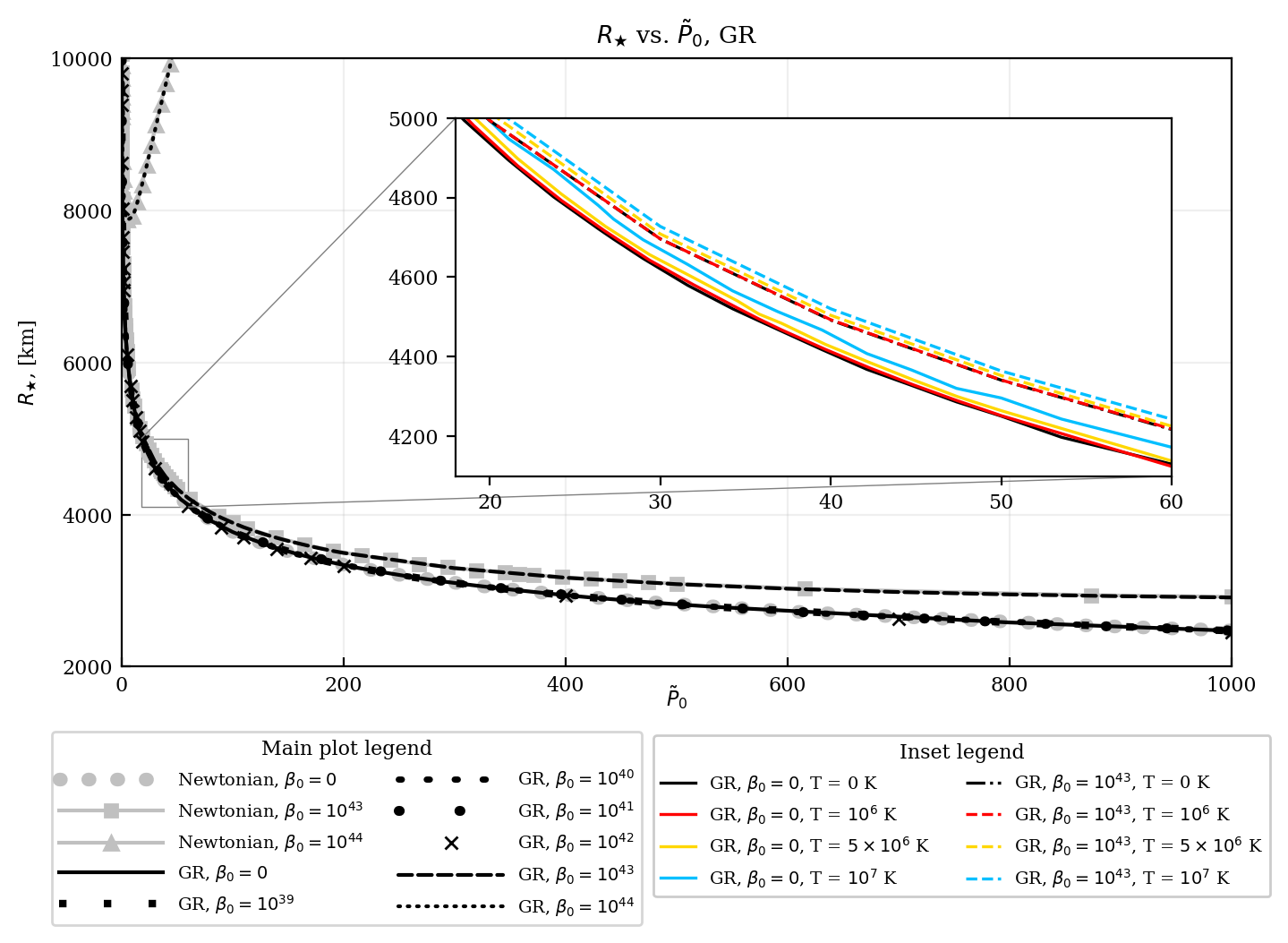}
    \caption{$R_{\star}$ vs. $\tilde{P}_0$ for GUP-modified GR white dwarfs at finite temperatures.}
    \label{fig:GR_rvP_QGUP}
\end{figure}

\begin{figure}[t!]
    \centering
    \includegraphics[width=1\textwidth]{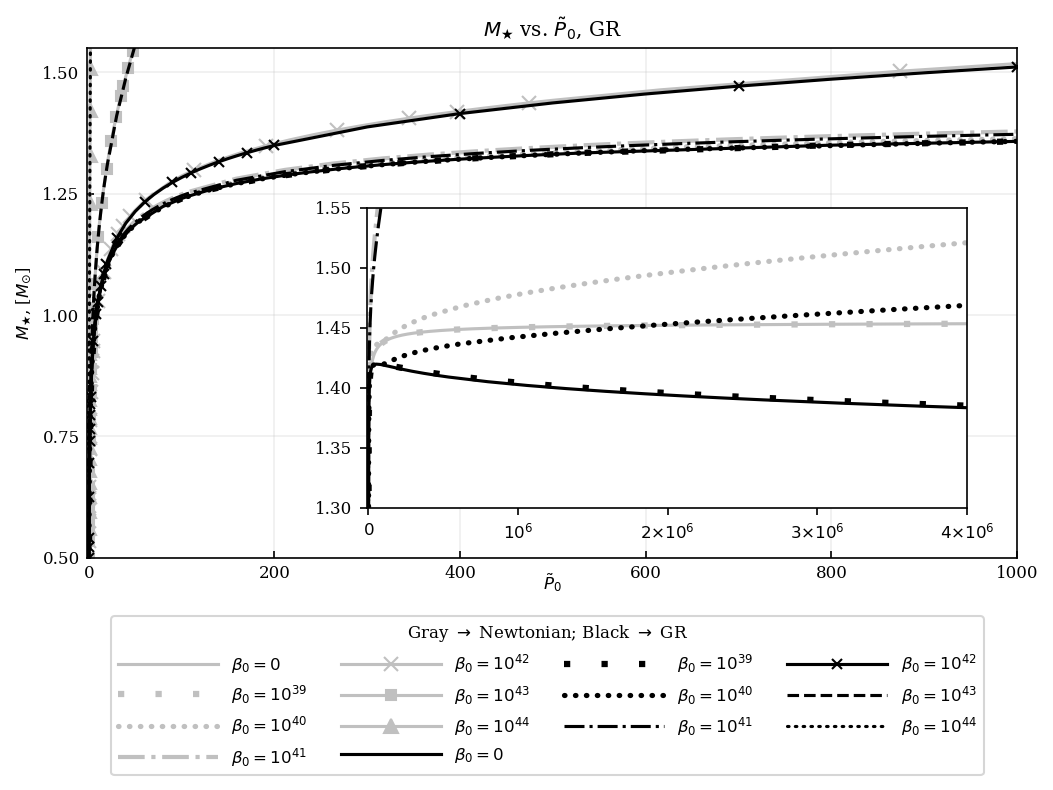}
    \caption{$M_{\star}$ vs. $\tilde{P}_0$ for GUP-modified GR white dwarfs at finite temperatures.}
    \label{fig:GR_MvP_QGUP}
\end{figure}

Here, we reproduce our solutions in the previous section, this time using the exact TOV equations. The GR stellar radii across central pressures on Figure \ref{fig:GR_rvP_QGUP} obtain similar values with the Newtonian radii, with thermal effects still occurring at small pressures (see Figure \ref{fig:NC_rvP_QGUP}). For the GR stellar masses on Figure \ref{fig:GR_MvP_QGUP}, prominent differences with the Newtonian masses are most evident for $\beta_0 = 10^{39}, 10^{40}$, as shown on the inset. The $\beta_0 = 10^{39}$ masses closely resemble the GR ideal masses -- both peak at $1.419M_{\odot}$, then subsequently decrease as $\tilde{P}_0$ increases. For small enough $\beta_0$, it appears that GR continues to support the white dwarf's gravitational collapse beyond a finite mass limit \cite{silbar2004neutron, shapiro2008black, glendenning, carvalho2018general}. At $\beta_0 = 10^{40}$ however, the masses begin to increase beyond the limit, the GR masses in particular falling below the Newtonian masses in dotted grey, because the onset of this increase occurs beyond the supposed GR mass dip. This behavior of a dip due to gravitational collapse being reversed by a sudden rise in mass is a clearer depiction of the``bouncing effect" mentioned in the Newtonian mass-radius relations.

\begin{figure}[t!]
    \centering
    \includegraphics[width=1\textwidth]{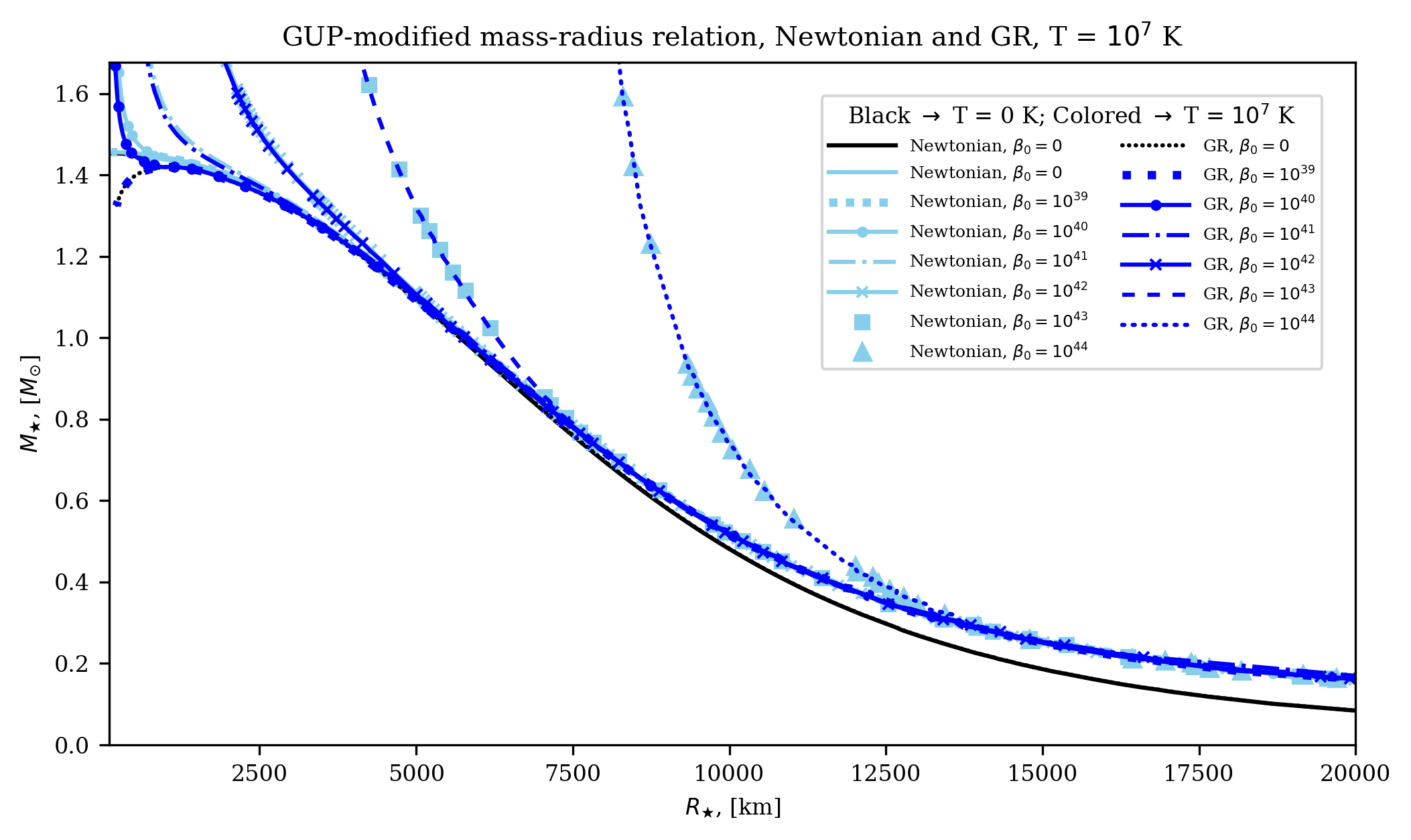}
    \caption{GUP-modified mass-radius relations for GR white dwarfs at T = $10^7$ K.}
    \label{fig:GR_mass-radius_qgup_t}
\end{figure}

Finally, we obtain the GR mass-radius relations on Figure \ref{fig:GR_mass-radius_qgup_t}. Behaviors found here are reminiscent of those found in the Newtonian relations, where the expected thermal deviation is found in the low-mass regime and the GUP effects occur in the high-mass regime. We find the GR tendency to produce slightly reduced masses occurring for $\beta_0 = 10^{39},10^{40}$, where relations are slightly shifted downward, but the tendency seems to disappear for larger $\beta_0$. For $\beta_0 = 10^{39}$ the relation follows the ideal GR case, having a mass limit $1.419M_{\odot}$ at the stellar radius $R_{\text{lim,GR}} = 1084$ km. For larger $\beta_0$, we still find the relations being lifted from the ideal case, shooting toward infinity as the modified white dwarfs acquire arbitrarily large masses and radii.

\begin{figure}[t!]
    \centering
    \includegraphics[width=1\textwidth]{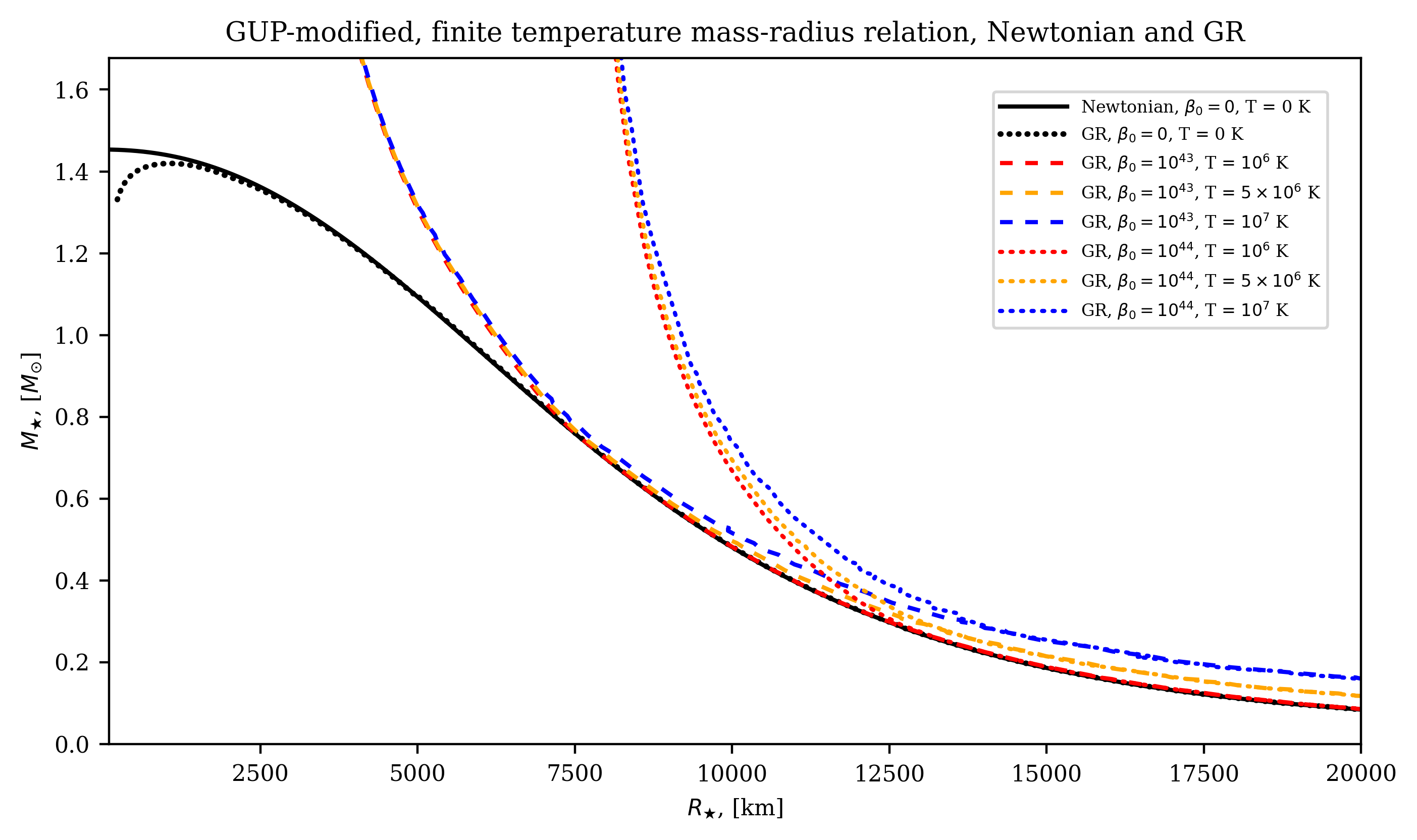}
    \caption{GUP-modified mass-radius relations for GR white dwarfs, for temperatures T = $10^6, 5 \times 10^6, 10^7$ K.}
    \label{fig:GR_mass-radius_qgup_comparison}
\end{figure}

It is also worth noting that for a large enough temperature and $\beta_0$, we obtain a mass-radius relation that is completely removed from the ideal cases, as shown on Figure \ref{fig:GR_mass-radius_qgup_comparison} for emphasis. It is here that we find the effects of temperature and GUP truly overlapping, as opposed to the exclusivity of these two types of deviations within the low and high mass regions when both parameters are relatively small. Although quite the extraordinary finding, we are still led by observation to believe that these relations are nonphysical, given the existence of heavier compact objects beyond the Chandrasekhar mass limit.

%%%%%%%%%%%%%%%%%%%%%%%%%%%%%%

	\section{Conclusions and Recommendations}
	\label{section: section_5}

In this study, we have explored the phenomenological effects of quantum gravity, as manifested by GUP modifications in finite temperature white dwarfs. This was done by applying the modified phase space volume to the thermodynamic properties of a degenerate Fermi gas, from which the equation of state was derived. The first approach in calculating this EoS involved expanding the EoS as a Taylor series by treating the GUP parameter as perturbative, which led to the derivation of general formulas to calculate the EoS up to any order of approximation in the series. In the low-pressure regime, this perturbative EoS exhibited the expected thermal deviation across all orders of approximation. At high pressures however, the EoS showed conflicting behaviors at each order, as dictated by the sign and magnitude of the last term of the truncated Taylor series. Furthermore, the perturbative approach proved to be numerically impractical for large orders. We therefore resorted to using the non-perturbative approach, where we found that the EoS should saturate toward constant energy densities in the high-pressure regime, which is the exact behavior shown by the modified zero temperature EoS. The non-perturbative EoS differed from this modified cold EoS in the thermal deviations present at low pressures.

The non-perturbative EoS was then used to complete the stellar structure equations. In Newtonian gravity, the solutions to these equations produced a mass-radius relation with two primary deviations from the ideal case: in the low-mass regime, white dwarfs obtain slightly larger masses, which is what is expected when temperature is involved, while in the high-mass regime, white dwarfs obtain arbitrarily large masses and radii, the same nonphysical result observed for modified cold white dwarfs. Furthermore, increasing the quadratic GUP parameter causes stronger deviations in the mass-radius relation, affecting more low-mass white dwarfs in the region near the Chandrasekhar limit. The solutions in GR follow the same behavior, albeit shifting the relations downward in the high-pressure regime. Indeed, GR continues to decrease the masses of white dwarfs approaching Chandrasekhar's limit, but in the context of the quadratic GUP, the modified white dwarfs do not evade unbounded growth in size and mass. We also saw that for a large enough temperature and GUP parameter, these two effects overlap, leading to mass-radius relations that are completely removed from the ideal case. 

A possible extension to this paper can be made by performing statistical analyses to determine which modified mass-radius relation best describes observational data of white dwarf parameters. Methods employed by Ref. \citen{bedard2017measurements} may be used, where one can also find a table of model-independent white dwarf masses and radii. Another option is the method employed by Ref. \citen{belfaqih2021white}, in which data was compared to mass-radius relations obtained using a linear-quadratic GUP.

Future studies may employ more sophisticated EoS's in conjuction with quadratic GUP and finite temperature considerations. One example is the Salpeter EoS, which considers the effects of local inhomogeneities of the electron distribution, or the relativistic Feynman-Metropolis-Teller EoS, which generalizes the Salpeter EoS by taking into account $\beta$-equilibrium and Coulombic interactions under a full relativistic fashion \cite{boshkayev, rotondo2011relativistic}. Additionally, more realistic mass-radius relations may be produced by also considering effects of the white dwarf's angular momentum, magnetic field, and lattice energy, which in combination with quantum gravity (as manifested by GUP) may produce a finite mass limit \cite{ong, moussa}.

As mentioned in the introduction, one may also choose to employ different approaches to the GUP, varied in derivation and form due to the absence of a full theory of quantum gravity \cite{garay1995quantum,wang,wang2012quantum}. An example is the linear-quadratic GUP used in Ref. \citen{belfaqih2021white}, whose GUP factor is of the form $(1-\beta p + \gamma p^2)^{-4}$. Various combinations of $\beta$ and $\gamma$ produce a corresponding variety of mass-radius relations, some with mass limits and others without, but a finite temperature extension in conjuction with these combinations is yet to be performed. Another approach is the extended GUP (EGUP) used in Ref. \citen{ong2018generalized}, which takes the form $\Delta x \Delta p \sim \hbar[1 + \alpha (\Delta p)^2 + \beta (\Delta x)^2]$, with the $\beta (\Delta x)^2$ term being related to the cosmological constant. Through heuristic methods, EGUP is shown to ``protect" the Chandrasekhar limit -- it would be interesting to confirm this prediction using the numerical methods of our study. To do this, one must first find the correct deformation of phase space when using the EGUP. As a benchmark, this has been found by Ref. \citen{chung2019extended} for the modified commutator $[ x,  p ] \sim \hbar[1 + \alpha x^2]$ known as the EUP (extended uncertainty principle), or the EGUP without the term proportional to $p^2$.

Lastly, one may choose to modify the TOV equations by taking into account the cosmological constant $\Lambda$. By assuming a nonzero $\Lambda$ in the Einstein field equations, Ref. \citen{silbar2004neutron} notes that the right-hand side of $dP/dr$ in the TOV equations is only extended by an extra term. A finite $\Lambda$ is associated with the presence of dark energy \cite{peebles2003cosmological}, and the acceleration of the universe's expansion. While its value is considered to be small, one might observe deviations in the mass-radius relations, hence completing this modified TOV equation with a finite temperature EoS, in combination with any of the previous suggestions, could produce some interesting results.

	\section*{Acknowledgments}

James Tuñacao thanks the Department of Science and Technology - Science Education Institute and the Iron Wood Corporation for funding this research project through their respective scholarship programs.
		
%\nocite{*}
\bibliographystyle{ws-ijmpd}
\bibliography{bibfile}

\end{document}